# Imaging electronic states on topological semimetals using scanning tunneling microscopy

András Gyenis[1], Hiroyuki Inoue[1], Sangjun Jeon[1], Brian B. Zhou[1], Benjamin E. Feldman[1], Zhijun Wang[1], Jian Li[1], Shan Jiang[2], Quinn D. Gibson[3], Satya K. Kushwaha[3], Jason W. Krizan[3], Ni Ni[2], Robert J. Cava[3], B. Andrei Bernevig[1], Ali Yazdani[1*]

[1]Joseph Henry Laboratories and Department of Physics, Princeton University, Princeton, NJ 08542, USA

[2]Department of Physics and Astronomy and California NanoSystems Institute, University of California at Los Angeles, Los Angeles, CA 90095, USA

[3]Department of Chemistry, Princeton University, Princeton, NJ 08542, USA

*Corresponding author. Email: yazdani@princeton.edu

**Following the intense studies on topological insulators, significant efforts have recently been devoted to the search for gapless topological systems. These materials not only broaden the topological classification of matter but also provide a condensed matter realization of various relativistic particles and phenomena previously discussed mainly in high energy physics. Weyl semimetals host massless, chiral, low-energy excitations in the bulk electronic band structure, whereas a symmetry protected pair of Weyl fermions gives rise to massless Dirac fermions. We employed scanning tunneling microscopy/spectroscopy to explore the behavior of electronic states both on the surface and in the bulk of topological semimetal phases. By mapping the quasiparticle interference and emerging Landau levels at high magnetic field in Dirac semimetals $Cd_3As_2$ and $Na_3Bi$, we observed extended Dirac-like bulk electronic bands. Quasiparticle interference imaged on Weyl semimetal TaAs demonstrated the predicted momentum dependent delocalization of Fermi arc surface states in the vicinity of the surface-projected Weyl nodes.**

## Introduction

The band structure of topological semimetals such as Dirac and Weyl semimetals arises from the discrete number of crossing points of the conduction and valence bands [1-11]. In analogy with relativistic quantum field theory [12–14], the crossing of two non-degenerate or two doubly degenerate bands leads to emergent low-energy, bulk excitations described as two-component Weyl fermions or four-component Dirac fermions, respectively.

In general, the hybridization of electronic bands results in avoided crossings and hinders the realization of such semimetals. The symmetries of the system, therefore, are crucial to restrict the phase space of interactions among the bands and stabilize the semimetal phases. In the presence of time-reversal and space-inversion symmetries, the electronic bands are doubly degenerate at every crystal momentum and a three-dimensional Dirac semimetal phase appears at the quantum critical point between trivial and topological insulators [3,15-19]. As such realization of Dirac nodes demands a fine-tuning of material parameters, e.g., the strength of spin-orbit interaction, it is difficult to obtain them in real materials. However, the hybridization of bands can be prevented by additional crystallographic symmetries such as discrete rotational symmetries, which lead to protected degeneracies of a three-dimensional Dirac node even in stoichiometric crystalline systems [20–23].

Breaking either the time-reversal or inversion symmetry [3-10] lifts the two-fold degeneracy of the bands and reduces the four-fold crossings to two-fold ones in the bulk band structure. In the vicinity of these Weyl points, the low-energy excitations are equivalent to massless Weyl fermions. Each Weyl node acts as a topological charge since it is a source or sink of the Berry flux depending on its chirality [4,24-27]. In contrast to Dirac semimetals, the Weyl nodes possess intrinsic topological stability: translationally invariant, small perturbations can only shift the position of Weyl points in the energy-momentum space. However, bringing a pair of opposite Weyl points to the same location may result in a gap opening. These topological charges in the bulk Brillouin zone can lead to protected gapless states on the surface in the form of disjointed Fermi arcs connecting the projected Weyl-nodes with opposite chirality. [22-23,28-29].

Dirac and Weyl semimetals have an unusual surface-bulk connectivity that sets them apart from other topological electronic systems. The existence of Weyl nodes leads to exciting physical properties such as the chiral anomaly [30-34], nonlocal transport [35-36], novel quantum oscillations [37-38], diverging diamagnetism [39] and unusual optical conductivity [40]. Recent photoemission measurements provided evidence of linearly dispersing bulk bands and the Fermi arc structure of the surface states both on Dirac [41-43] and Weyl semimetals [44-51]. Here we employed scanning tunneling microscopy (STM) to probe the exotic electronic structure of topological semimetals with high energy and spatial resolution, which could shed light on unique aspects of these systems [52-56].

**Mapping the band structure with STM**

A powerful method to probe the electronic band structure of materials with STM is to use Fourier-transform scanning tunneling spectroscopy (FT-STS) [57-60]. When quasiparticles originating from the surface and/or bulk states of the system scatter from potential barriers caused by lattice imperfections, the incoming and outgoing electronic waves with crystal momenta of $k_i$ and $k_f$ form a quantum interference pattern with $q = k_i - k_f$ characteristic wavevector. The quasiparticle interference (QPI) results in modulation in the local density of states of the sample, which can be mapped locally by measuring the conductance between the STM tip and the sample surface. The Fourier transform of the obtained conductance map (FT-STS map) can be approximately characterized by wavevectors connecting regions with high density of states in the band structure.

More precisely, if $\psi_{n\alpha}(k)$ denotes the Bloch eigenstate of the *n*-th Bloch band with spin component $\alpha$, energy of $\varepsilon_n(k)$ and lifetime $\Gamma$, we can introduce the spectral function matrix:

$$A_{\alpha\beta}(k,\omega) = -\frac{1}{\pi}\sum_n \text{Im}\left(\frac{1}{\omega-\varepsilon_n(k)+i\Gamma}\right)\psi_{n\alpha}(k)\psi_{n\beta}^+(k).$$

The total and spin spectral density can be obtained as $\rho_i(k,\omega) = Tr[\sigma_i A(k,\omega)]$ for $i = 0$, and $i = 1, 2, 3$ respectively, where $\sigma_i$ are the Pauli matrices. In the case that all scattering events are allowed, the FT-STS map can be approximated by the joint-density of states:

$$JDOS(q,\omega) = \int dk\ \rho_0(k,\omega)\rho_0(k+q,\omega).$$

Taking into account the effect of the spin texture in the scattering process, we can describe the resulting FT-STS map with the spin-selective scattering probability:

$$SSP(q,\omega) = \frac{1}{2}\sum_{i=0}^{3}\int dk\ \rho_i(k,\omega)\rho_i(k+q,\omega).$$

In the pure state limit, this expression reduces to a more commonly used form:

$$SSP(q,\omega) = \int dk\ \rho_0(k,\omega)T(k,q)\rho_0(k+q,\omega),$$

where $T(k,q) = |\langle S(k)|S(k+q)\rangle|^2$ and $S(k)$ is the spin of the band at momentum $k$.

When calculating the integrals above, the integration boundary in the momentum space has to be chosen carefully. Here, we follow a recent theoretical work [61], and we introduce a form factor which envelopes the spectral density $\tilde{\rho}_i(k,\omega) = \rho_i(k,\omega)\times\exp(-k^2/\xi^2)$ and calculate the integrals over multiple, higher order Brillouin zones without a sharp cutoff at the boundary of the first Brillouin zone.

To enhance the signal-to-noise ratio of the FT-STS maps, generally, a series of data analysis steps are applied to the experimentally obtained real space conductance maps. In this paragraph, we briefly summarize the most common techniques used for this purpose. Firstly, we note that although an impurity is a source of the QPI signal, the conductance value measured on top of it is dominated by tunneling into its bound states rather than by the signal originating from the QPI. Therefore, it is beneficial to replace the real space conductance values obtained at the position of the impurities with the mean conductance value of the map at the given energy to suppress this effect. Drift due to piezo motion or thermal expansion during the long time scale (typically several days) of acquisition of a conductance map also decreases the sharpness of the QPI features. To compensate this artifact, we use a previously developed drift correction method on the real space images [62]. Also, the noise of the scanning piezo can be reduced by normalizing the measured conductance value with the recorded tunneling current at each point of the map. Further improvement of the QPI map can be achieved by suppressing the edge effects arising from the finite size of the conductance map, which can be done by applying certain window functions (e. g. tukey window function). Finally, the Fourier transform of the maps are symmetrized based on the symmetry of the underlying lattice. We note that the reported features are apparent without any of these data analysis steps.

In addition to the FT-STS technique, investigating the energy structure of Landau levels formed in high magnetic fields can provide another experimental approach to characterize the electronic band structure. Previously, Landau level scanning tunneling spectroscopy has been utilized to characterize the energy momentum dispersion in two-dimensional systems through the emerging quantized energy levels [63-66]. In three-dimensional systems, however, the presence of magnetic field leads to the formation of Landau bands dispersing along the momenta parallel to the applied field. In an STM experiment, the van-Hove singularities of these Landau bands cause peaks in the measured projected bulk density of states. The $\varepsilon_N$ Landau level energies for a linearly dispersing band is expected to have $\varepsilon_N = \varepsilon_{DP} \pm v_F\sqrt{2eB(N+\gamma)/\hbar}$ scaling behavior, where $\varepsilon_{DP}$ is the Dirac energy, $v_F$ is the Fermi velocity, $e$ is the electron charge, $B$ is the magnetic field, $N$ is the orbital index, $\gamma$ is the phase offset of quantum oscillations, $\hbar$ is the Planck's constant divided by $2\pi$. Hence, the square-root dependence of the singularities in the energy spectrum signals the linear dispersion relation with the slope corresponding to the Dirac velocity.

**STM on Dirac semimetals, Cd$_3$As$_2$ and Na$_3$Bi**

The first theoretically proposed [22-23] and experimentally identified [41-43] three-dimensional Dirac semimetals are Na$_3$Bi and Cd$_3$As$_2$. Both materials are characterized by strong spin-orbit interaction and by an inverted band structure. First principle calculations [22-23] showed that the band inversion occurs in the vicinity of the $\Gamma$ point between the Cd - 5$s$ and As - 4$p$ states in Cd$_3$As$_2$, while in Na$_3$Bi, it occurs between Na - 3$s$ and Bi - 6$p_{x,y}$ orbitals. The pair of Dirac points in both materials is located along the $k_z$ direction in the first Brillouin zone and stabilized by C$_4$ and C$_3$ rotational symmetry of the crystal structure (Fig. 1(a) and Fig. 2(a)) in case of Cd$_3$As$_2$ and Na$_3$Bi, respectively.

Fig. 1(b) and Fig. 2(b) show atomically flat topographic images of the natural (112) cleaved surface of Cd$_3$As$_2$ [52] and the (100) surface of the first batch of Na$_3$Bi(1) [53]. The (112) surface of Cd$_3$As$_2$ displays a pseudo-hexagonal structure corresponding to the As layer, whereas the Na$_3$Bi(1) surface displays stripe-like topographic modulations along the $c$ axis of the crystal. The FT-STS measurements carried out on these surfaces display long wavelength ripples in the real space density of states (Fig. 1(d) - (g) and Fig. 2(d) - (g)). The Fourier transform of the spectroscopic maps (Fig. 1(h) - (k) and Fig. 2(h) - (k)) can be characterized by a rather isotropic set of *q* scattering wavevectors in both materials. By determining the boundary of the QPI signal in the FT-STS maps as a function of energy, the scattering vectors exhibit an approximately linear energy-evolution even extending to energies far from the Dirac point (Fig. 1(c) and Fig. 2(c)). At low-energies, the diverging quasiparticle scattering wavelength and sample inhomogeneity become comparable, and therefore prevents resolving of the band dispersion close to the Dirac point by FT-STS.

When high magnetic field is applied perpendicular to the (112) face of Cd$_3$As$_2$ and the (001) surface of the second batch of Na$_3$Bi(2) [52-53], the measured spectra exhibit a series of pronounced peaks in the tunneling spectra as illustrated in Fig. 3(a) and (c), respectively. In the case of Cd$_3$As$_2$, each LL splits into two peaks in a non-uniform way indicating that the spin degeneracy is lifted by the combination of orbital and Zeeman splitting. By mapping the energy of the peaks as a function of magnetic field, we find that they correspond to electron-like Landau levels. The position of the Landau level peaks exhibits an $\varepsilon_N \propto \sqrt{(N+\gamma)B}$ scaling, providing evidence of Dirac-like dispersion of the bands (Fig. 3(b) and (d)). For Cd$_3$As$_2$, the Dirac energy is located slightly below - 200 meV ($v_F \approx 6.2$ eVÅ), while in case of Na$_3$Bi(2) it is located closer to the Fermi energy at - 20 meV ($v_F \approx 5.1$ eVÅ). We also note that model calculation for Cd$_3$As$_2$ showed that the band inversion is expected to be as small as 20 meV [52], which prevents us to grab the signatures of Lifshitz transition in this material.

Since studying the physics of Weyl fermions requires either broken inversion or time-reversal symmetry, application of magnetic field, in principle, is one of the possible routes to reach the Weyl semimetal phase from the Dirac phase. In case of $Cd_3As_2$, the normal vector of the natural cleavage plane is tilted away from the rotational symmetry axis, hence, applying magnetic field perpendicular to the plane eliminates the Weyl nodes [52]. In case of $Na_3Bi$, the high spatial variability of the measured spectra and surface disorder prevented access to the physics of Weyl fermions [53].

**STM on Weyl semimetals TaAs**

The first experimentally confirmed, inversion-symmetry-breaking Weyl semimetal compounds were found in the transition metal monopnictide materials class [27-28, 44-51]. We here focus on TaAs, which possesses a body-centered tetragonal lattice with four adjacent Ta and As sublayers forming a unit cell (Fig. 4(a)). The subsequent sublattices are shifted by a fraction of the unit cell, hence the inversion symmetry is naturally broken in the crystal. Previous theoretical calculations [27-28] showed that the bulk band structure has 12 pairs of Weyl points located close to the mirror planes. Four pairs of the Weyl nodes (W1) sit on the $k_z = 0$ plane, while the remaining 8 pairs (W2) are located away from it. The projection of the Weyl nodes on the (001) surface, however, results only in 8 pairs of the Weyl points in the surface BZ (Fig. 4(b)) because the pairs of the W2 nodes have the same in-plane momentum.

Similar to our measurements on the Dirac semimetals [52-53], we utilized quantum interference technique to explore the physics of Weyl fermions in TaAs [54]. Fig. 4(c) shows the (001) cleaved, As-terminated surface of single crystal TaAs with a few different crystallographic defects, which give rise to strong modulations in the surface density of states due to scattering of quasiparticles (Fig. 4(d)). The Fourier transform of this conductance map reveals that the interference pattern can be characterized by a rich variety of scattering wavevectors arising from the rather complex surface band structure (Fig. 4(e)). We calculated the SSP maps (Fig. 4(f)) based on our density-functional theory (DFT) results, which unlike in the case of the quasiparticle interference measured in Dirac semimetals, is undoubtedly insufficient to describe the measured QPI: the SSP displays a large set of wave vectors, which is absent in the measured data. Understanding the discrepancy between the observed and predicted QPI signal, however, reveals an important new aspect of the physics of electrons in a Weyl semimetal: the momentum dependent delocalization of the surface Fermi arcs into the bulk.

Since Fermi arcs connect projected bulk Weyl nodes with opposite chirality, the connectivity of the surface states with the bulk states is a unique aspect of Weyl semimetals. A remarkable consequence of the surface-bulk connectivity of topological semimetals has been possibly illustrated in recent magnetotransport measurements [37-38]. In contrast to conventional materials, where the electron cyclotron orbit consists of either surface or bulk trajectories on the Fermi surfaces perpendicular to the magnetic field, topological semimetals feature unusual, closed magnetic orbits consisting of the combination of the two. When an electron travels on a disjoint Fermi arc and reaches the end point of the arc (the projection of the Weyl node), it loses its surface characteristic and traverses through the bulk until it reaches the other surface. This unusual quantum path involving electrons sinking around the Weyl nodes leads to anomalous quantum oscillations in transport measurements. In our STM measurements [54], we revealed how this "sinking effect" influences the QPI pattern on the surface of a Weyl semimetal.

In order to quantitatively understand the surface-bulk connectivity of TaAs, we first study the atomic characteristics of both the bulk and surface states as a function of their momenta using our DFT calculation. Focusing on the bulk properties, we calculate the contribution of the Ta - $5d$ and As - $4p$ orbitals on the bulk bands. As an example, Fig. 5(a) shows a linecut in the bulk band structure through the W1 and W2 Weyl points, where the red and green colors indicate the corresponding Ta and As weights, respectively. At momenta far from the Weyl points, the As orbitals dominate the wavefunctions, while the Weyl nodes themselves are clearly monopolized by the Ta orbitals (at W1 and W2 the ratio of the contribution of Ta/As orbitals is about 6:1 and 14:1, respectively).

In the case of the surface states, we can determine the momentum-dependent atomic character of the Fermi arcs by projecting the DFT calculated spectral density on the top Ta and As layers separately and subtracting the two projections from each other (Fig. 5(b)). We observe that the bowtie-shaped features are more pronounced on the As layer (green), while the spoon-like features are more visible on the Ta layer (red). Thus, we can see that bowties are generally associated with As orbitals and spoons with Ta orbitals. This behavior is also expected from the momentum-space location of the states: spoons are located closer to the projected Weyl cones, which are dominated by the Ta contribution, and therefore they have a stronger connection to the bulk. Additionally, when we follow the intensity of the bowtie and spoon features as a function of layer position, it is apparent that the contribution of the bowties decreases more rapidly, signaling a weaker connectivity to the bulk as compared to the spoon features (Fig. 5(c)). A similar trend can be observed by looking at the total spectral weight on the As and Ta layers (Fig. 5(d)).

To summarize our atomic orbital characterization of the Weyl states, we see that the spoon features associated with Ta have stronger bulk character, while the bowtie surface states residing more strongly on As have a more surface-like character. As the QPI observed in an STM measurement is naturally dominated by the electronic states which remain on the surface, in order to understand our data we need to focus on states which have a smaller probability to connect to the bulk. Therefore, we introduce a weighted Fermi surface [54] which consists of states projected only on the As sites instead of using the projection onto the entire unit cell. As Fig. 5(e) shows, this projection causes the spoon features to become more faint in the weighted Fermi surface but leaves the bowtie features mostly unaffected.

There are three sets of scattering wavevectors in the first Brillouin zone: between bowties and bowties (*P* vectors), between bowties and spoons (*Q* vectors) and finally between spoons and spoons (*R* vectors) with their intensity rapidly decreasing from *P* to *R* owing to the intensity difference between bowties and spoons in the weighted Fermi surface. The possible *P*, Q and *R* scattering vectors are shown separately in Fig. 6(a) – (c), and the corresponding QPI signal in the calculated SSP is highlighted by red and blue rectangles in Fig. 6(d) – (f). In the case of the *P* and *Q* vectors, the same QPI vectors can be clearly identified in the measured FT-STS maps (Fig. 6(g)-(h)), illustrating an excellent match between theory and measurements. Although some of the *S* vectors can be found in the SSP calculation (Fig. 6(f)), their weak intensity or their overlap with other wave vectors prevents us from detecting them in our measurement (Fig. 6(i)).

The strong connection between the atomic orbital character and the Fermi surface features suggests that the different lattice sites carry different parts of the QPI signal. To investigate this idea, we decompose the measured conductance maps according to Ta and As sublattices. By taking the real space conductance map and considering the measured signal only on top of Ta or As sites (associated with local hills and valleys in the topographic image as shown in Fig. 6(l)), two maps can be constructed: the Ta-site and the As-site selected conductance map. The Fourier transform of each map remarkably shows that they can be independently described by distinct set of scattering vectors corresponding to the *P* and *Q* vectors (Fig. 6(j)-(k)). While the As-site selected conductance map can be understood as scatterings between bowties, the quasiparticle interference measured on the Ta site can be described as scatterings between bowties and spoons.

The revealed momentum-dependent surface-bulk connectivity is insensitive to the topological nature of the surface states: both topological and trivial states are delocalized into the bulk at momenta close to the projection of the Weyl points. The presented DFT calculation, however, can be used to access additional information about the topological classification of the surface states on TaAs [54]. The calculated band structure encircling W1 and W2 (on a trajectory indicated on Fig. 7(a)) is shown on Fig. 7(b) and Fig. 7(c), respectively. In case of the W1 node, the gray shaded regions corresponds to the bulk, whereas yellow and blue lines correspond to the topological surface states connecting the conduction and valence band. The energy-momentum diagram reveals that the surface states give rise to four crossings at the Fermi energy and consequently all four surface states around W1 have topologically nontrivial origin. The number of Fermi arcs naively contradicts the charge of W1, which is two. However, by taking into account the "chirality" of the states (the sign of their velocities when they cross the Fermi energy), we can see that the three chiral arcs (crossings 2, 3, 4) and the one anti-chiral (crossing 1) arc result in a net number of two crossings in agreement with the total topological charge of W1. We also note that one of the topological arcs is possibly hybridized with a trivial surface state, therefore, the distinction between topological or trivial states is rather vague in this case. The classification of surface states around W2 is clearer. As Fig. 7(c) shows, the black line corresponds to a trivial, while the green line is a topological surface state. Therefore, the topologically protected Fermi arc state can be associated with the state directly connecting the two W2 points, while the inner bowtie has trivial character.

Topological semimetals represent a novel quantum phase of matter, where touching of bulk electronic bands leads to non-trivial band topology. Our scanning tunneling microscopy measurements [52-54] with high energy and spatial resolution provided a crucial insight into the physics of these systems by mapping the electronic behavior in high magnetic field and visualizing the scattering of both the bulk and surface carriers establishing a topological connectivity between those states.

**Figure captions**

**Figure 1.** (a) Crystal structure of $Cd_3As_2$ ($a$ = 12.67 Å, $c$ = 25.48 Å), with the adjacent gray plane indicating the cleaved (112) surface direction. (b) STM topographic image of a 145 Å x 145 Å large, (112) surface of $Cd_3As_2$ ($V_{bias}$ = -250 mV, $I_{set}$ = 500 pA). The nearest neighbor atomic distance is approximately 4.4 Å, which matches the As-As or Cd-Cd distance in the (112) plane. (c) The size of the measured QPI disk on the FT-STS maps as a function of energy. The differently colored points refer to different QPI bands. The extracted Fermi velocity is: 6.15 ± 0.86 eVÅ. (d)-(g) Real space conductance maps ($V_{bias}$ = 600 mV, $I_{set}$ = 300 pA, $V_{osc}$ = 7 mV) acquired at various energies, and (h)-(k) the corresponding Fourier transform of the images. $G$ is the measured conductance, $G_0$ is the mean and $\sigma$ is the standard deviation of the conductance. PSD stands for power spectral density. Data is from [52].

**Figure 2.** (a) Crystal structure of $Na_3Bi$ ($a$ = 5.45 Å, $c$ = 9.66 Å), with the adjacent gray plane indicating the cleaved (100) surface direction. (b) STM topographic image of a 145 Å x 145 Å large, (100) surface of $Na_3Bi$(1) ($V_{bias}$ = -500 mV, $I_{set}$ = 30 pA). The spacing of the stripe like modulation is approximately 9.1 Å, which is close to the lattice constant in the $c$ direction. (c) The size of the the measured QPI disk on the FT-STS maps as a function of energy. The extracted Fermi velocity is: 4.34 ± 1.26 eVÅ. (d)-(g) Real space conductance maps ($V_{bias}$ = 600 mV, $I_{set}$ = 500 pA, $V_{osc}$ = 5 mV) acquired at various energies, and (h)-(k) the corresponding Fourier transform of the images. Data is from [53].

**Figure 3.** (a) Landau levels measured on (112) $Cd_3As_2$ ($V_{bias}$ = 300 mV, $I_{set}$ = 500 pA, $V_{osc}$ = 1.2 mV) showing split, doublet structure. The curves are shifted for clarity. (b) The extracted energy-momentum dispersion from the Landau level singularities with the $k$ values (light red dots with error bars) obtained from the QPI measurements (Fig. 1(c)). (c) Landau levels measured on the (001) surface of $Na_3Bi(2)$ ($V_{bias}$ = -200 mV, $I_{set}$ = 100 pA, $V_{osc}$ = 5 mV), and (d) the corresponding dispersion of the Landau level singularities. For comparison, we included the Landau level and extracted QPI dispersion (light red dots with error bars) obtained on the (100) surface of $Na_3Bi(1)$. Data is from [52-53].

**Figure 4.** (a) Crystal structure of TaAs ($a$ = 3.44 Å, $c$ = 11.66 Å). The unit cell contains 4 pairs of Ta and As layers. (b) DFT-calculated Fermi-arc structure of the (001) surface on TaAs. Arrows indicate the in-plane spin direction, red and green dots show the position of the projected Weyl nodes with positive and negative chirality, respectively. (c) Topographic image of TaAs surface ($V_{bias}$ = 240 mV, $I_{set}$ = 120 pA) and (d) the measured conductance map at 40 meV on the same area ($V_{bias}$ = 240 mV, $I_{set}$ = 120 pA, $V_{osc}$ = 5 mV). (e) Fourier transform of the conductance map. (f) Calculated SSP based on the Fermi surface shown in (b). Data and calculations are from [54].

**Figure 5.** (a) The bulk band structure along the line (shown in the inset) crossing the W1 and W2 Weyl points. The ratio of the red and green colored disks corresponds to the Ta/As orbital weight ratio at the given state. (b) Difference between the spectral density projected to the top As and top Ta layer. (c) Intensity of the spectral density separately integrated over the bowtie and spoon features as a function of layer index. (d) Full Brillouin zone integrated spectral density on the As and Ta layers as a function of layer index. (e) Calculated weighted Fermi surface.

**Figure 6.** Scattering vectors on the WFS between (a) bowtie and bowtie, (b) bowtie and spoon and (c) spoon and spoon features. (d)-(f) Calculated SSP with rectangles indicating the corresponding quasiparticle interference vectors. (g)-(i) Measured FT-STS maps with the same superimposed rectangles as in (d)-(f). (j) As and (k) Ta site selective FT-STS maps. (l) Topographic image showing the position of Ta and As sites.

**Figure 7.** (a) Topological nature of the surface states close to the Y point. Blue, yellow and green colors indicate the topological Fermi arcs, whereas black color corresponds to the trivial states. (b)-(c) The calculated band structure along two closed paths enclosing W1 and W2 nodes in the momentum-space shown in (a). The gray regions corresponds to the bulk bands, while the surface bands localized at the As termination are indicated by red dots. The color of the surface bands are the same as in (a), and the numbers indicate the Fermi-level crossings. Calculation is from [54].

**Acknowledgement**

Work at Princeton was supported by Army Research Office-Multidisciplinary University Research Initiative (ARO-MURI) program on topological insulators W911NF-12-1-0461, Gordon and Betty Moore Foundation as part of the Emergent Phenomena in Quantum Systems (EPiQS) initiative (GBMF4530), by NSF Materials Resarch Science and Engineering Centers (MRSEC) programs through the Princeton Center for Complex Materials DMR-1420541, NSF-DMR-1104612, NSF-DMR-0819860, NSF CAREER DMR-0952428, NSF EAGER Award NOA - AWD1004957, Department of Energy DE-FG-02-05ER46200, DE-SC0016239, Packard Foundation, Simons Investigator Award, Schmidt Fund for Innovative Research and Keck Foundation. This project was also made possible through use of the facilities at Princeton Nanoscale Microscopy Laboratory supported by grants through ARO-W911NF-11-1-0379, ARO-W911NF-1-0262, ONR-N00014-14-1-0330, ONR-N00014-13-10661, ONR- N00014-11-1-0635, US. Department of Energy-Basic Energy Sciences (DOE-BES), Defense Advanced Research Projects Agency – U.S.Space and Naval Warfare Systems Command (DARPA-SPWAR) Meso program N6601-11- 1-4110, LPS and ARO-W911NF-1-0606, and Eric and Wendy Schmidt Transformative Technology Fund at Princeton. Work at University of California-Los Angeles was supported by the DOE-BES (DE-SC0011978).

Figure 1

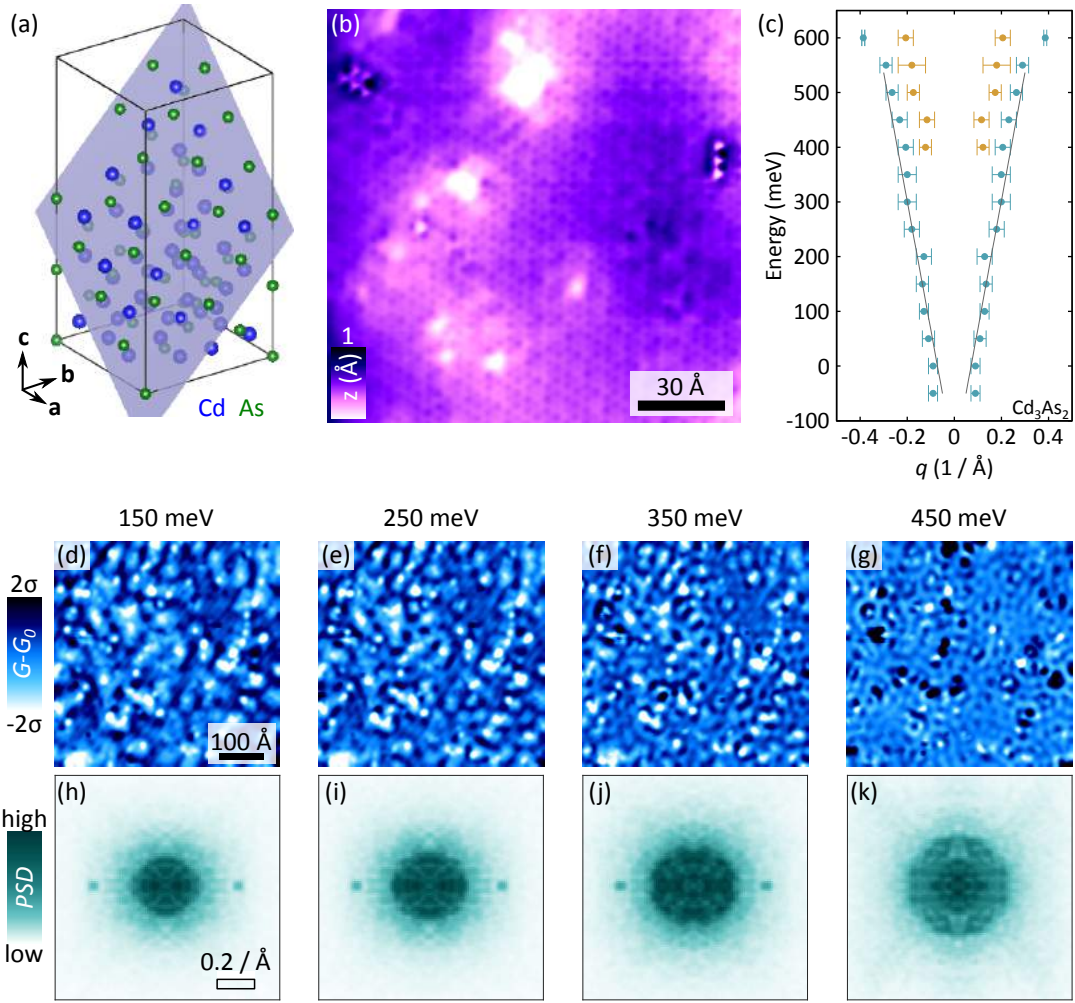

Figure 2

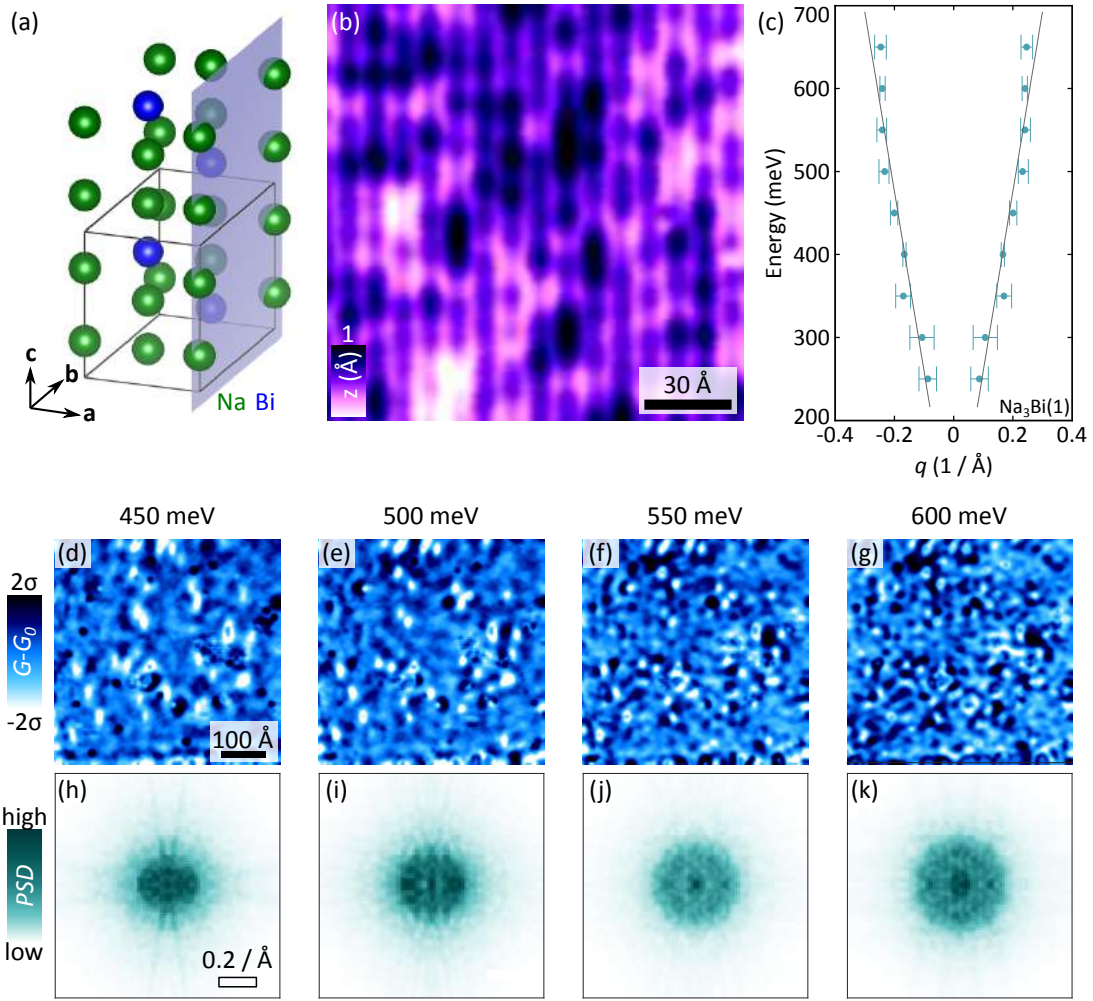

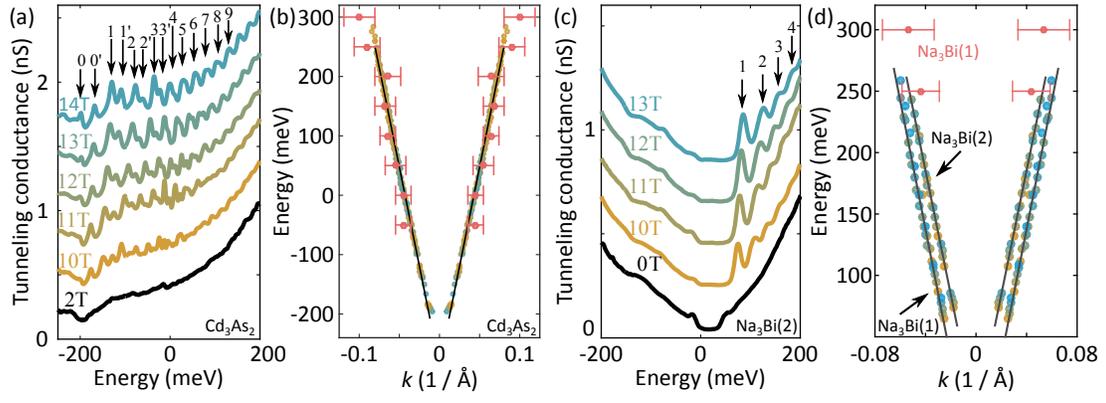

Figure 4

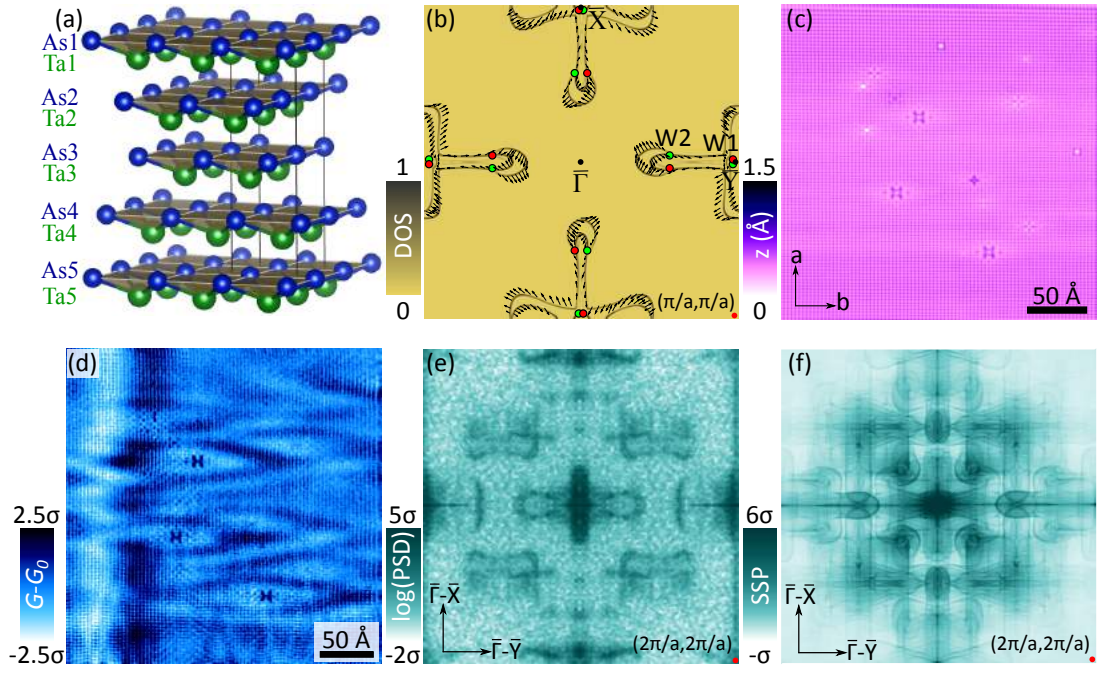

Figure 5

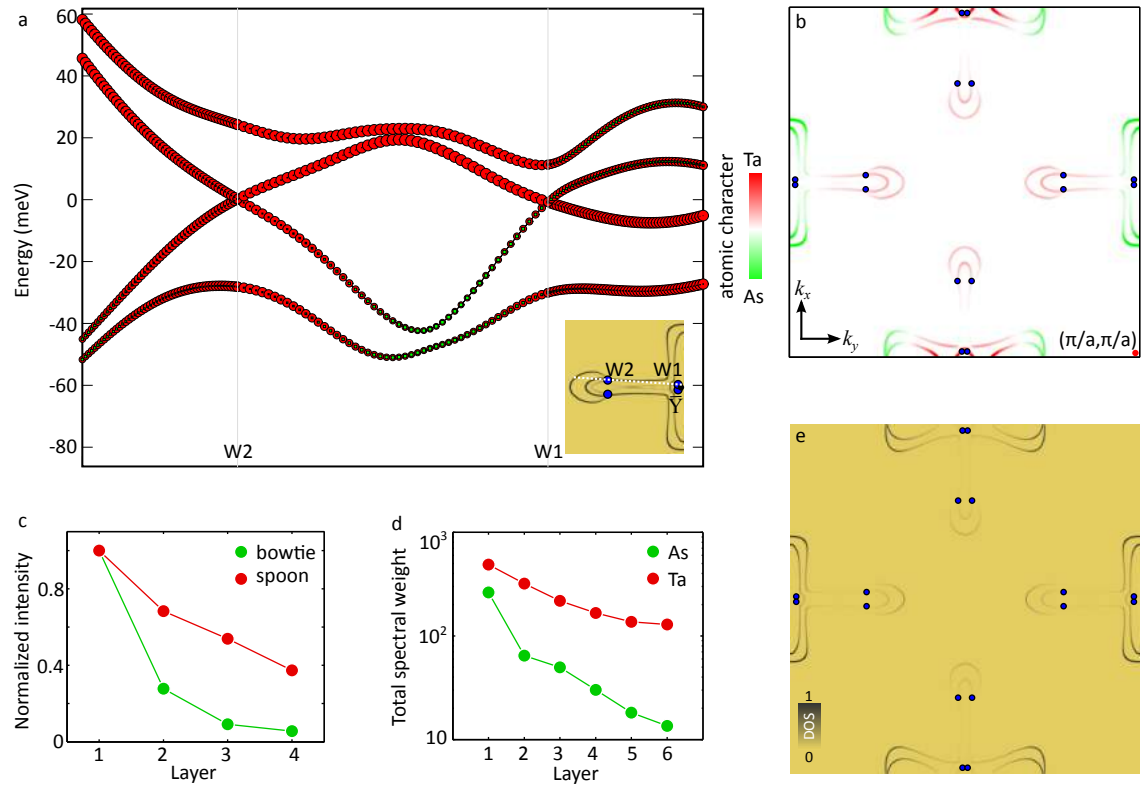

Figure 6

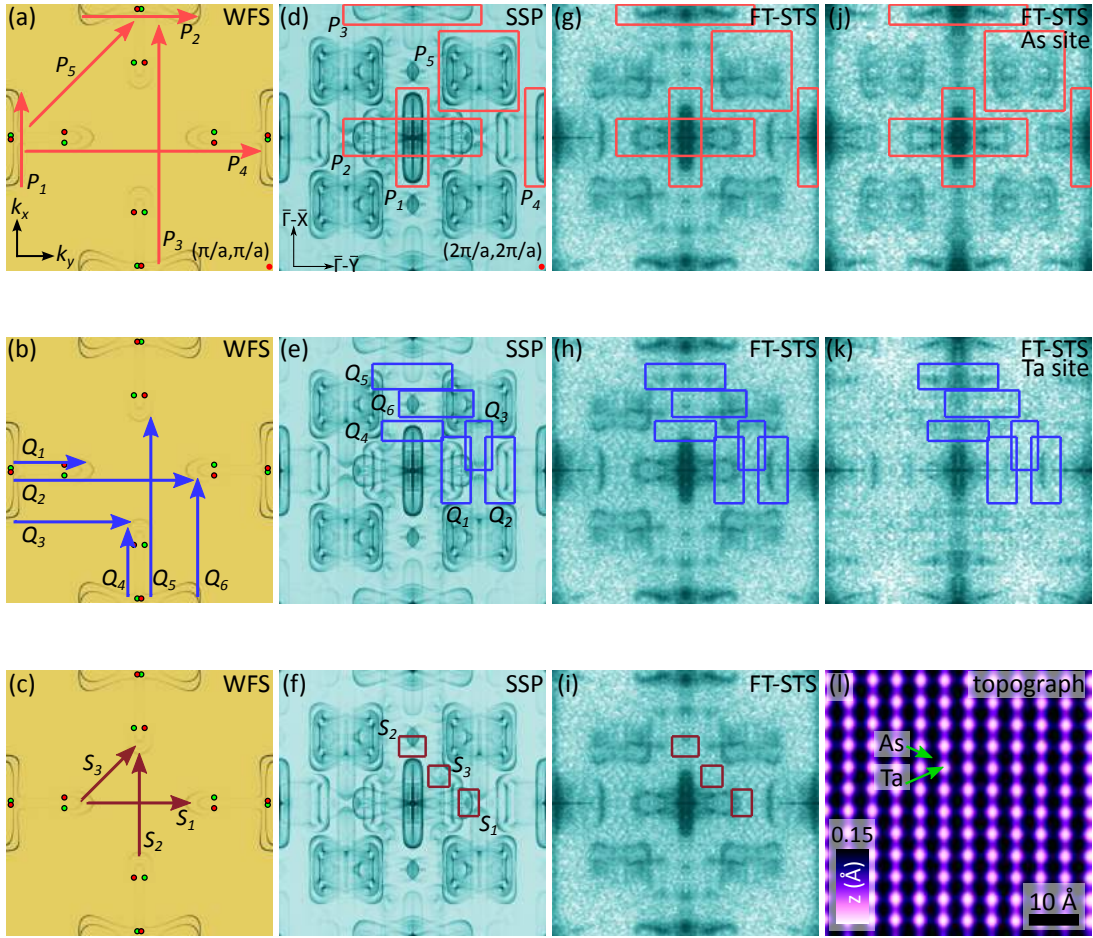

Figure 7

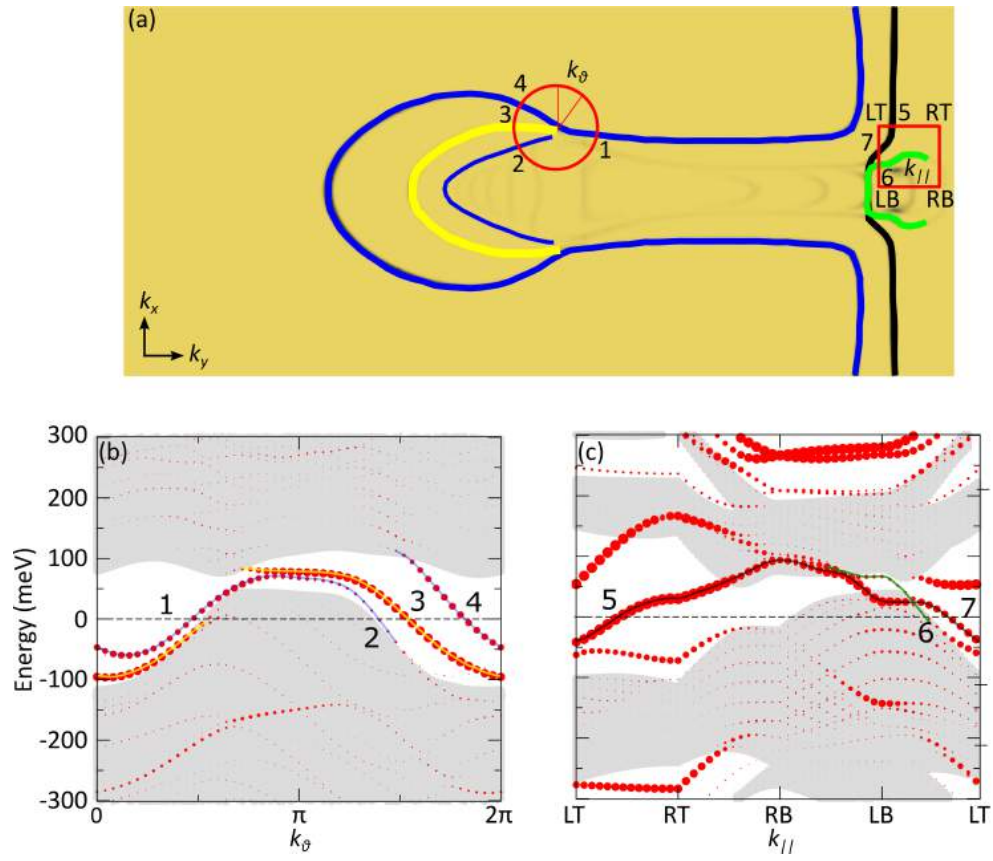